\documentstyle[referee]{mn}
\newcommand{\reference}{\bibitem}
\def\beq{\begin{equation}}
\def\eeq{\end{equation}}
\def\bey{\begin{eqnarray}}
\def\eey{\end{eqnarray}}
\def\beqarray{\begin{eqnarray}}
\def\eeqarray{\end{eqnarray}}

\def\mpc{\,{\rm {Mpc}}}
\def\Mpc{\,{\rm {Mpc}}}

\def\kpch{\,{h^{-1}{\rm kpc}}}
\def\mpch{\,h^{-1}{\rm {Mpc}}}
\def\kms{\,{\rm {km\, s^{-1}}}}

\def\v200{V_{200}}

\def\md{m_{\rm g}}
\def\Onow{\Omega_{\rm m,0}}

\def\my{{\rm M_\odot yr^{-1}}}

\def\Mgas{M_{\rm g}}
\def\Mdot{\dot{M}_{\star}}
\def\vc{V_{\rm c}}
\def\tstar{t_\star}
\def\th{t_{\rm H}}
\def\fB{f_{\rm B}}
\def\sigmav{\sigma_{\rm v}}
\def\lambdaeff{\lambda_{\rm eff}}
\def\Gyr{\,{\rm Gyr}}
\def\L850{L_{850\rm \mu m}}
\input epsf

\title[]
{The Host Haloes of Lyman Break Galaxies and Sub-millimeter
Sources}

\author[]
{Chenggang Shu$^{1,2,4}$, Shude Mao$^{3}$, H.J. Mo$^{2}$
\thanks {E-mail: cgshu@center.shao.ac.cn,
smao@jb.man.ac.uk,
hom@mpa-garching.mpg.de} \\
$^1$ Shanghai Astronomical Observatory, Chinese Academy of
Sciences, Shanghai 200030, China\\
$^2$ Max-Planck-Institut f\"ur Astrophysik,
      Karl-Schwarzschild-Strasse 1, Postfach 1317
      D-85741 Garching, Germany \\
$^3$ Jodrell Bank Observatory, Univ. of Manchester,
	Macclesfield, Cheshire SK11 9DL, UK\\
$^4$ The Joint Lab of Optical Astronomy, Chinese Academy of
Sciences}
\date{Accepted ........
      Received .......;
      in original form ......}
\pubyear{1999}

\begin{document}
\maketitle
\begin{abstract}
We use the observed sizes and star formation rates
of Lyman-break galaxies (LBGs) to understand their properties
expected in the hierarchical scenario of galaxy formation.
The observed size distribution
constrains the masses (or circular velocities) of the host
haloes of LBGs from below, because small haloes can only 
host galaxies with small sizes. The observed star-formation
distribution constrains the masses from above, because
the number density of massive haloes in duty cycle
is too low to host all the observed galaxies. Assuming a flat 
CDM model with a cosmological constant
($\Omega_\Lambda=0.7$), we find that consistency with observations
requires the circular velocities of LBG host haloes 
to be in the range $100$--$300\,{\rm km\,s^{-1}}$.  
The star formation in these haloes is quite efficient;
a large fraction of all the gas in them can form stars on a time scale
of about 10--20 per cent of the Hubble time at redshift of 3.
The predicted comoving correlation length 
of these objects is $\sim 3 h^{-1}{\rm Mpc}$, and the 
predicted velocity dispersion of their stellar   
contents is typically $70\,{\rm km\,s^{-1}}$.
It is likely that these LBGs are the progenitors of
galactic bulges and faint ellipticals. 

 The same prescription applied to larger haloes in 
the CDM cosmogony predicts the existence of galaxies 
with star formation rates 
$\sim 1000\,{\rm M_\odot\,yr^{-1}}$ at redshift $z\sim 3$.
We explore the possibility of identifying these 
galaxies to be the bright sub-millimeter (sub-mm) sources
detected by SCUBA. The model predicts that the host haloes
of these sub-mm sources are massive, with typical 
circular velocity $\sim 350\, {\rm km\,s^{-1}}$. 
The typical star formation timescale in these systems
is about 10 per cent of the Hubble time at redshift 3, and
the comoving number density of galaxies (in their duty 
cycle) is $10^{-5}$--$10^{-4}h^3 {\rm Mpc^{-3}}$.
These galaxies are predicted to be strongly correlated, 
with a comoving correlation length of $\sim 7 h^{-1}{\rm Mpc}$.
They are also predicted to be strongly correlated with LBG population 
at the same redshift, with a cross-correlation 
length $(4-5)\,h^{-1}{\rm Mpc}$.
The descendants of the bright sub-mm sources should reside 
in clusters of galaxies at the present time, and it is likely
that these objects are the progenitors of giant 
ellipticals. We estimate that about 15 per cent of
the observed sub-mm background is contributed by this population
of objects in the redshift range $2.5$--$3.5$;
a comparable contribution is made by LBGs in the same redshift 
range. The model predicts the existence of a relatively bright and red
population of galaxies at $z\sim 3$, which may be observed in the K-band.

\end{abstract}

\begin{keywords}
galaxies: formation - galaxies: structure - galaxies: halo -
galaxies: starburst - galaxies: elliptical
\end{keywords}

\section {Introduction}

 At present there are two main populations of objects 
which may be crucial to our understanding of the formation of
bright galaxies in the universe.
The first is the Lyman-break population
at redshifts $z\sim 3$, observed by the Lyman break technique 
(Steidel, Pettini \& Hamilton 1995). The second is 
the population detected by some recent sub-mm detectors
such as SCUBA on JCMT (Holland et al. 1999; Blain et al. 
1999). Galaxies in both populations are found to
show vigorous star formation activity, 
and so a large fraction of stars we observe today may have 
been formed in these systems  

 The Lyman-break technique is quite efficient in 
identifying star forming galaxies at $z\sim 3$. 
At the moment, about 1000 galaxies have been selected
by this technique
and confirmed by followup spectroscopic observations
(Steidel et al. 1999b). The sample is selected according to 
a flux limit in the optical $R$-band ($R_{\rm AB}\la 25.5$)
which corresponds to UV in the rest-frame. 
Thus, the selection may imply a lower limit on the 
star formation rates in the selected galaxies. 
The typical star formation rate 
is $\sim 50 {\rm M_\odot\,yr^{-1}}$ (all values are quoted
for a flat universe with $\Omega_{\rm m,0}=0.3$,
$\Omega_\Lambda=0.7$ and $H_0=70\,{\rm km\,s^{-1}\,Mpc^{-1}}$), 
assuming a dust correction of a factor of about 5 
(e.g. Steidel et al. 1999a,b; Pettini et al. 1999a,b;
Adelberger \& Steidel 2000). The comoving number density
of the Lyman-break galaxies (hereafter LBGs)
is quite high, $\sim 2\times 
10^{-3}h^3{\rm Mpc^{-3}}$ (Adelberger et al. 1998),
comparable to the number density of local bright galaxies.
They are also strongly clustered in space,  with a correlation 
length comparable to that of  normal galaxies in the local universe
(Adelberger et al. 1998; Giavalisco \& Dickinson 2000).
With a typical half-light radius of about 
$1.5\,h^{-1}{\rm kpc}$ (Giavalisco et al. 1996; Lowenthal et
al. 1997), LBGs are quite small compared to the
normal galaxies in the local universe.

 Since the Lyman-break technique selects galaxies according to 
their rest-UV luminosity, it may miss dust-enshrouded
star-forming galaxies, because the UV photons from young 
stars in such galaxies may be absorbed by dust and 
re-emitted in the far-infrared. Such galaxies can be detected
in the sub-mm band, and so the recent discovery
of strong sub-mm sources by SCUBA on JCMT may provide important 
clues to the star formation at high redshifts (see Blain et al. 1999). 
Unfortunately, current observations are still unable to determine 
the nature of these sub-mm sources. In particular, 
spectroscopic data on these sources are generally 
difficult to obtain, and so most of them do not have
measured redshifts. If these strong sub-mm
sources have redshifts similar to the LBG
population ($z\sim 3$), as many people believe,  
the star formation rates in them may be as high as 
$\sim 1000\,{\rm M_\odot\,{\rm yr}^{-1}}$.
If this is the case, the bright sub-mm sources may be 
an extension of the LBG population to higher star formation rates. 
Observational support for this comes from the
fact that galaxies with the highest UV star formation 
rates are indeed detected statistically in the SCUBA map
(Peacock et al. 2000).

Theoretically, the nature of the LBG population has been
studied by many authors, but the results are still
inconclusive. There have been suggestions that LBGs 
might be associated with massive dark haloes at $z\sim 3$
(e.g. Mo \& Fukugita 1996; Baugh et al. 1998).
Such association allows one to make 
specific predictions for the clustering of the LBG 
population, and the model predictions are generally 
consistent with the observed high correlation amplitude 
(Mo \& Fukugita 1996; Baugh et al. 1998; Jing \& Suto 1998;
Moscardini et al. 1998;
Haehnelt, Natarajan \& Rees 1998; Peacock et al. 1998;
Wechsler et al. 1998; Coles et al. 1998; Mo, Mao \& White 1999; 
Arnouts et al. 1999; Kauffmann et al. 1999; 
Katz, Hernquist \& Weinberg 1999). The star formation properties  
and structure of the LBG population have also been 
studied under this hypothesis (e.g. Baugh
et al. 1999;  
Mo, Mao \& White 1999; Shu 2000). However, because of the 
uncertainties in the observations, the
theoretical interpretation is not unique. For example, 
some authors suggest that LBGs are starbursts triggered
by galaxy interactions (Kolatt et al. 1999; 
Somerville et al. 2000; Wechsler et al. 2000).
In this scenario, the freedom in model construction
is larger. For example, in one version, 
all LBGs are assumed to be starbursts triggered by 
galaxy-galaxy collisions, and so the star 
formation rate is not closely related to the mass 
of the host halo (Kolatt et al. 1999). 
Alternatively, LBGs may correspond to massive 
systems which were undergoing temporary mergers
(either major or minor), as suggested by
Somerville et al. (2000). 

 Theoretical interpretation of the sub-mm sources 
is even more uncertain. 
At the moment, the observational data are still too 
sparse to rule out any speculations. 
It is possible that these sources belong to the extension 
of the LBG population to higher star formation rates, 
as mentioned above, but current observations cannot rule
out the possibility that they form a 
completely distinct population. It is therefore 
important to make model predictions and outline key future 
observations which can be used 
to distinguish different scenarios.

In this paper, we use recent observational results to set 
constraints on  theoretical models for the LBG population.
Our approach is different from those in earlier analyses
in that it is largely empirical.
We show that important constraints can already be obtained
from current observational data.
We use models motivated by the existing observational results
to make further predictions which can be tested with future observations.
In particular, we examine the properties of bright
sub-mm sources in the same theoretical framework, and make
predictions for this largely unexplored population. 

The layout of this paper is as follows. We 
describe our empirical approach in Section 2. 
Predictions for the LBG population and for the sub-mm sources
are presented in Sections 3 and 4, respectively.
We summarize and make further discussions of our results in Section 5.

Since we are dealing with observations at high redshifts,
a cosmological model has to be assumed in order to 
transform observables into physical quantities.
Throughout this paper we assume a flat model 
with matter density $\Onow=0.3$ and a cosmological 
constant corresponding to $\Omega_\Lambda=0.7$. 
We write the Hubble constant as 
$H_0=100h\,{\rm kms^{-1}Mpc^{-1}}$.

\section{An Empirical Approach to the Lyman-break Population} 
\subsection{Observational Inputs}

There are two important observational constraints on the properties
of the LBG population. The first is their size distribution
and the second is their star formation rates.
Based on the deep imaging data from the 
{\it Hubble Space Telescope},
the effective radii have been determined for 45 LBGs
brighter than $R=25.5$
(Giavalisco et al. 1996; Lowenthal et al. 1997).
In Figure 1 we plot the observed size distribution,
together with a log-normal fit of the data.
The number of LBGs with determined sizes is 
still quite small, and so significant revision of the
observational size distribution is possible in the future.

\begin{figure}
\epsfysize=9.5cm
\centerline{\epsfbox{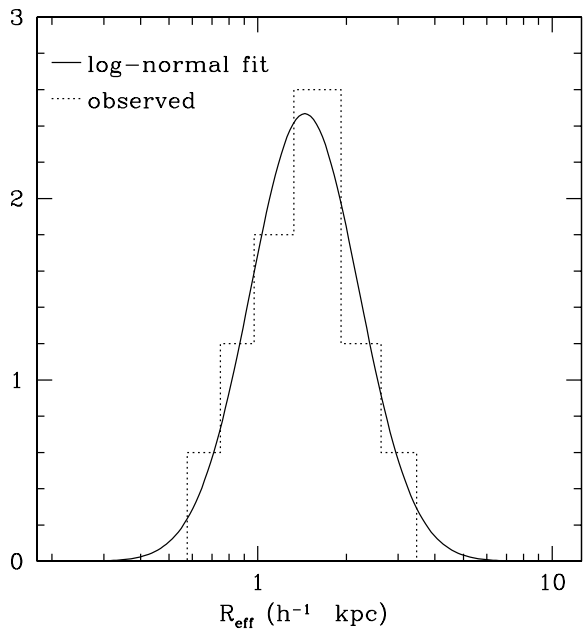}}
\caption{The histogram shows the observed distribution of 
the half-light radius of LBGs obtained from the data 
of Giavalisco et al. (1996) and Lowenthal et al. (1997)
using a flat cosmogy with $\Onow=0.3$, $\Omega_\Lambda=0.7$.
The solid curve is a log-normal fit to the data.}
\end{figure}

 The UV luminosity function of the LBG population 
is straightforward to derive from the observed magnitudes
(Steidel et al. 1999a,b) and the result is shown in Figure 2.
However, in order to transform the UV luminosity function
into a distribution in star formation rate, a correction 
for dust extinction has to be made. This turns out to be
difficult, because the correction depends on the
details of the dust distribution in individual galaxies. 
Recently, Adelberger \& Steidel (2000) carried out a detailed 
analysis of dust extinction in LBGs, and their 
dust-corrected luminosity function is shown in Figure 2.  
The average dust correction is a factor of 5, compared 
to the factor of 2 to 3 obtained earlier 
(Steidel et al. 1998; Dickinson 1998).
In our analysis, we will adopt the dust correction 
given by Adelberger \& Steidel (2000). We use only LBGs with 
observed $R_{AB}$ magnitude brighter than 25.5. 
For comparison, we also show results for two  other  
dust corrections: a factor of 3 (Dickinson 1998) and 
a factor of 11 (Cowie \& Barger 1999).
For convenience, we will refer these models of 
dust correlation as
\begin{itemize}
\item{Case (a)}: a dust correction of a factor of 3;
\item{Case (b)}: the dust correction based on Adelberger \& Steidel
(2000);
\item{Case (c)}: a dust correction of a factor of 11.
\end{itemize}
Once a dust model is adopted, the observed UV luminosity function  
can easily be converted into a distribution of 
the star formation rate. We use the model
as given in Steidel et al. (1999b).

\begin{figure}
\epsfysize=9.5cm
\centerline{\epsfbox{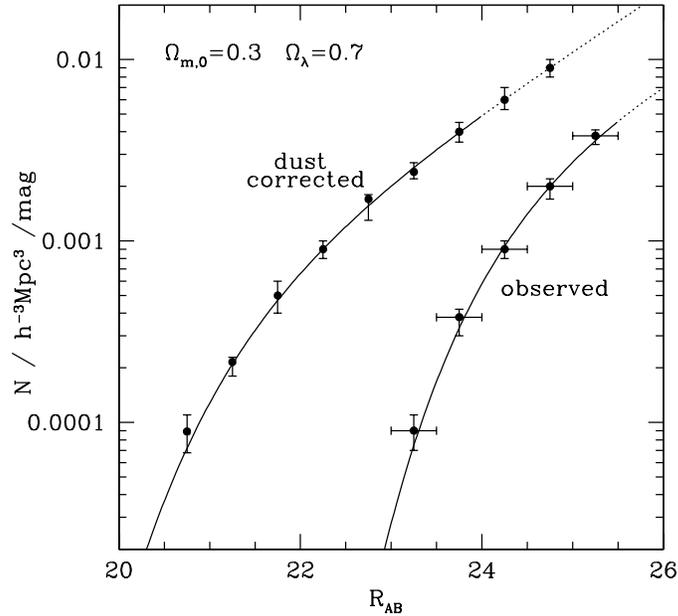}}
\caption{The observed and dust-corrected luminosity 
functions for Lyman-break galaxies, adopted 
from Adelberger \& Steidel (2000). 
The solid curves are fits to the data.
A cosmology with $\Onow=0.3$ and $\Omega_\Lambda=0.7$
is used in the calculation.}
\end{figure}

\subsection{The Empirical Approach}

 Given the size and star formation rate of a galaxy, 
we use the following procedure to predict the other properties
of the galaxy. We assume that the star formation in LBGs 
obeys a law similar to the Schmidt relation 
for local star-forming galaxies:
\beq\label{SFR_law}
\Sigma_{\rm SFR} = 2.5\times 10^{-10} 
\left({{\Sigma_{\rm gas}} \over
{M_{\odot}{\rm pc^{-2}}}}\right)^{1.4} M_{\odot}{\rm yr^{-1}}
{\rm pc^{-2}}\, ,
\eeq
where $\Sigma_{\rm SFR}$ is the star formation rate
(SFR) per unit area
and $\Sigma_{\rm gas}$ is the surface density of cold gas
(Kennicutt 1998). This empirical relation
is found to hold reasonably well for a variety 
of star-forming galaxies: from quiescent disks
to intensive starbursts. Although 
the validity of this relation for high-redshift star-forming 
galaxies has not yet been checked by observations, there is no 
obvious reason why star formation at $z\sim 3$ should
proceed in a dramatically different way. 
Assuming a surface-density profile for the 
gas, we can infer the total amount of cold gas
in a galaxy from its radius and star formation rate.
For an exponential surface-density profile
(which we adopt in our discussion), the total mass 
in cold gas ($M_{\rm g}$) is related to the 
star formation rate ${\dot M}_\star$ and effective 
radius $R_{\rm eff}$ (within which 
half of the UV luminosity is contained) as  
\beq\label{MSFR} \label{Mg}
\left( {M_{\rm g}} \over {10^{11}M_\odot}\right) \approx 
0.9 \times 10^{-2}\left({\dot M_\star}
\over{M_\odot{\rm yr^{-1}}}\right)^{0.71} 
\left({R_{\rm eff}}\over {\rm kpc}\right)^{0.57}\,.
\eeq
Notice that this relation is not very sensitive to the
density profile assumed, and so is a direct result 
of the empirical star formation relation. If we assume the
mass in the cold gas to be $m_{\rm g}$ times the mass of the 
dark halo, then the halo mass is 
\beq
M_{\rm h}={M_{\rm g}\over m_{\rm g}}\,.
\eeq
The value of $m_{\rm g}$ is not known. If all gas in 
a galaxy-sized halo settles into the centre as 
cold gas, then $m_{\rm g}=\Omega_{{\rm B}}/\Onow$
(where $\Omega_{{\rm B}}$ is the density parameter of
baryons), which is about $0.1$ if we take the cosmic 
nucleosynthesis value for a cosmology with 
$\Onow=0.3$ and $h=0.7$. At $z=3$, gas cooling is 
effective in galaxy-sized haloes, and so we expect
$m_{\rm g}\sim 0.1$ unless there is a process which can keep 
the gas from cooling. In our discussion we will treat 
$m_{\rm g}$ as a free parameter.  

 The observed flat rotation curves of spiral galaxies
suggest that the density profiles of dark haloes
are approximately isothermal at large radii. If we define the mass of 
a dark halo to be that within a radius 
$r_{\rm h}$ such that the mean density of the halo
is about 200 times the critical density (as given
by spherical collapse model), then the circular velocity
and radius of the halo can be inferred from   
\beq \label{vc}
V_c=\left[ 10 G H(z) M_{\rm h}\right]^{1/3},
~~~~~~~
r_{\rm h}=V_c/[10H(z)]\,,
\eeq
where $H(z)$ is the Hubble's constant at redshift $z$.
For a flat universe with $\Onow=0.3$, $H(z)/H_0\sim 4$ at
$z=3$. The concentration of a galaxy in its halo is defined to be
\beq\label{concentration}
\lambdaeff={\sqrt{2} R_{\rm d}\over r_{\rm h}}\,,
\eeq
where $R_{\rm d}= R_{\rm eff}/1.68$ is the disk scalelength
(Mao, Mo \& White 1998).
Defined in this way, $\lambdaeff$ resembles the spin parameter
of disk material in dark haloes (see Mo, Mao \& White 1998).
Hereafter we refer $\lambdaeff$ as the compactness parameter.
  
Based on the above discussion, we can construct a Monte Carlo 
sample of LBGs by randomly drawing $R_{\rm eff}$
and ${\dot M}_\star$ from their distribution functions. 
An extra assumption involved here is that 
${\dot M}_\star$ and $R_{\rm eff}$ are not strongly correlated.
This assumption is consistent with the observational data
for galaxies with $R_{\rm eff}$-measurements. Our conclusion does not
change significantly if a positive correlation 
between ${\dot M}_\star$ and $R_{\rm eff}$ 
($R_{\rm eff}\propto M_\star^{0.3}$, 
as expected if the compactness parameter is not correlated with
the halo circular velocity) is included.
  
\section{Predicted properties for the LBG Population}

\begin{figure}
\epsfysize=9.5cm
\centerline{\epsfbox{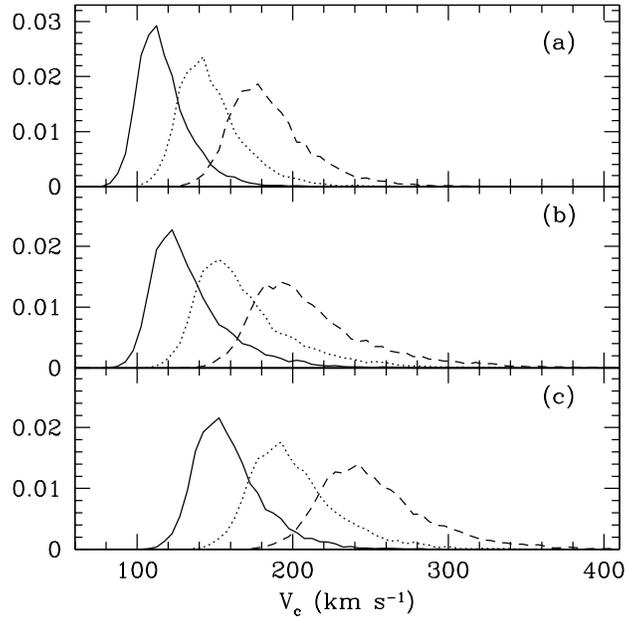}}
\caption{Predicted distribution for the circular velocity
$\vc$ of halos hosting LBGs. Panels (a)-(c) correspond to
the dust correction prescriptions of (a)-(c) defined in section 2.1.
For each dust correction model, the solid, dotted and dashed
lines are for $\md=0.12$, 0.06 and 0.03, respectively.}
\end{figure}

As discussed above, our empirical approach enables us to construct
a Monte-Carlo sample of LBGs that satisfy the observed distributions
of size and star formation rate. From this Monte-Carlo sample, we can
obtain various other properties of the LBG population and analyze
their distributions.

\subsection{Properties of the host haloes}

In the model outlined above, LBGs are housed in the central parts of
dark matter haloes. An important characteristic of the LBG population 
is the typical mass (or circular velocity) of their host haloes.
Figure 3 shows the predicted distribution of $\vc$
for the host halos of LBGs; three panels (a), (b) and (c) correspond
to the three prescriptions of dust correction as defined in
Section 2.1. For each dust correction model, 
the solid, dotted and dashed lines denote $\md$ of 0.12, 0.06 and 0.03,
respectively. One can see that as $\md$ increases, the
underlying host halo circular velocities decrease, as expected from
Equations (2-4). Figure 3 also indicates that for 
fixed $\md$, the required circular velocities of LBG hosts 
increase as the dust correction increases from case (a) 
to case (c). This trend can be easily understood as follows: as the dust
correction increases, a given observed UV luminosity corresponds to a higher
intrinsic total SFR, and to produce the higher SFR, the circular velocities
of underlying haloes have to increase.
More quantatively, for $\md=0.03$, the median values of $\vc$ are (180,
205, 250)$\kms$ for the three dust corrections cases (a, b, c),
respectively. For $\md=0.06$ and $0.12$, the corresponding
values for $\vc$ are (140, 165, 190)$\kms$,
and (110, 130, 160)$\kms$, respectively. In our favored
dust-correction model (b), the predicted median circular 
velocity is different from that obtained from early
analyses. The reasons for this difference will be discussed 
in more detail in Section 5.   

\begin{figure}
\epsfysize=9.5cm
\centerline{\epsfbox{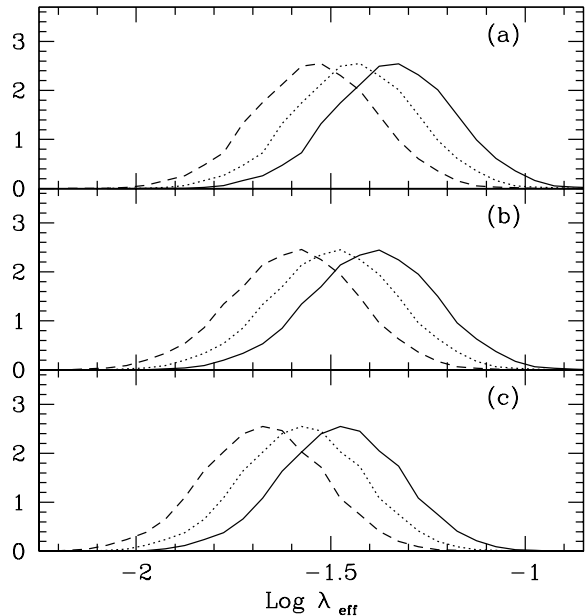}}
\caption{Predicted distributions of the effective spin
parameter $\lambdaeff$ for the host galaxies of
LBGs. The symbols have the same meaning as in Figure 2.} 
\end{figure}

Figure 4 shows the predicted distribution of the
compactness parameter as defined in equation (5).
For the dust correction case (b), the median values for
$\lambdaeff$ are 0.024, 0.028 and 0.033 for $\md=0.03, 0.06$ and 0.12,
respectively. For a given disk mass fraction $\md$, the 
predicted LBGs are more compact (i.e., with smaller $\lambdaeff$)
as the dust correction increases, because only the more
compact objects can produce high (intrinsic) SFR
while keeping the size distribution unchanged.  
For a given dust correction, the LBGs are more compact 
for smaller $\md$. This is because the predicted SFR drops as $\md$
decreases, so in order to simultaneously satisfy
the observed SFR and size distributions,
the objects have to become more compact.

The $\lambdaeff$-distributions for different dust corrections
and $\md$-choices all have log-normal shapes with almost identical
dispersion $\sigma_{\ln\lambdaeff}\approx 0.4$. This dispersion
is consistent with that found in numerical simulations 
for the spin parameter of dark matter haloes (Warren et al. 1992; 
Cole \& Lacey 1996; Lemson \& Kauffmann 1998). 
However, we would like to
emphasize that the $\lambdaeff$ parameter is related to but
not equivalent to the conventional spin parameter that describes
the angular momentum of dark haloes. Even if the initial specific angular
momentum of the gas is equal to that of the dark matter,
the gas may lose angular momentum to the dark halo in the 
settling process. Further, the self-gravity of the disk and the dark
matter response to baryonic settling in general reduce the disk size
and hence decrease the $\lambdaeff$ parameter (cf. Equation
5; 
see also Mo, Mao \& White 1998).
Also we caution that the gas settling process
is not well understood, so the final stellar component may
not be completely supported by rotation if the disk gravity
becomes important first; this may be the case for many of our systems
since they have $\md>\lambdaeff$.  In such cases, we expect that both
the angular momentum and the self-gravity of the gas are 
important in determining the final size. Indeed, if a LBG
is not supported by angular momentum, then the
compactness parameter is merely a measure of the concentration 
of the star-forming gas in a dark halo.

\begin{figure}
\epsfysize=9.5cm
\centerline{\epsfbox{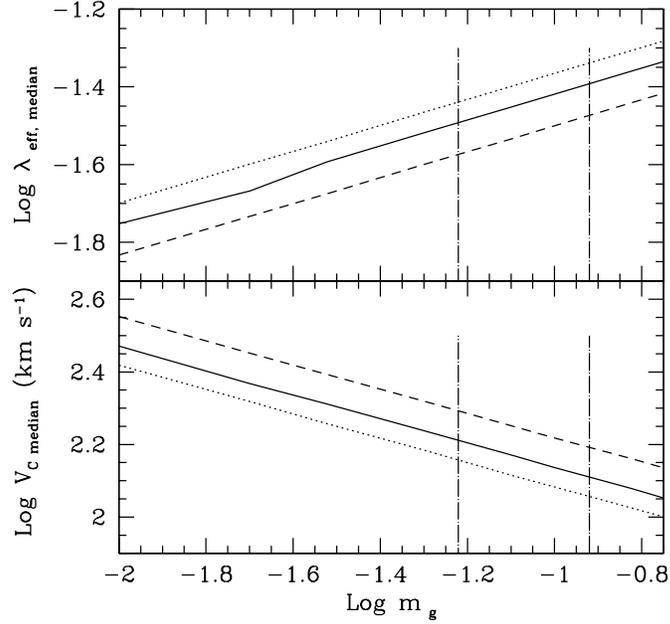}}
\caption{The predicted distributions of the median values of
$\vc$ and $\lambdaeff$ as a function of $\md$. The dotted, solid and
dashed lines correspond to the dust corrections of case (a), (b) and
(c), respectively. The two vertical dot-dashed lines bracket the reasonable
range of $\md$ from 0.06 and 0.12.
}
\end{figure}

Figure 5 shows the median values for $\vc$ and
$\lambdaeff$ as a function of $\md$. The
dotted, solid and dashed lines are for 
the dust corrections of case (a), (b) and (c), respectively. 
For all the dust corrections,
the median value of $\vc$ decreases with increasing $\md$ while
the $\lambdaeff$-dependence on $\md$ has an opposite sense. Again 
these trends are a result of the fact that we have taken the empirical
distributions for the SFRs and sizes of the LBG population
(see Equations 2-4).

While in Figure 5 we have plotted the parameter $\md$ over a wide
range, this parameter
is in fact limited by two considerations. First, it must be
smaller than or equal to the overall baryon mass fraction,
$\fB=\Omega_B/\Onow$. For the $\Lambda$CDM cosmogony with
$\Onow=0.3$, $\fB=0.12$, if
we adopt $\Omega_B=0.019h^{-2}$ as found by 
Burles \& Tytler (1998) from the deuterium abundance 
in QSO absorption line systems.
Second, as we will show in the next two subsections, the duty cycle of
LBGs and the abundance of dark haloes require
$\md\ga 0.06$. Hence, the typical value of $\md$ should 
be in the range 0.06 to 0.12.  These two constraints are shown as
two vertical dot-dashed lines in Figure 5. The range of $\md$ 
implies that the median value of the circular velocity $\vc$ for LBGs should be
 between 130 and 165$\,\kms$.

\subsection{Star formation timescale and duty cycle}

\begin{figure}
\epsfysize=9.5cm
\centerline{\epsfbox{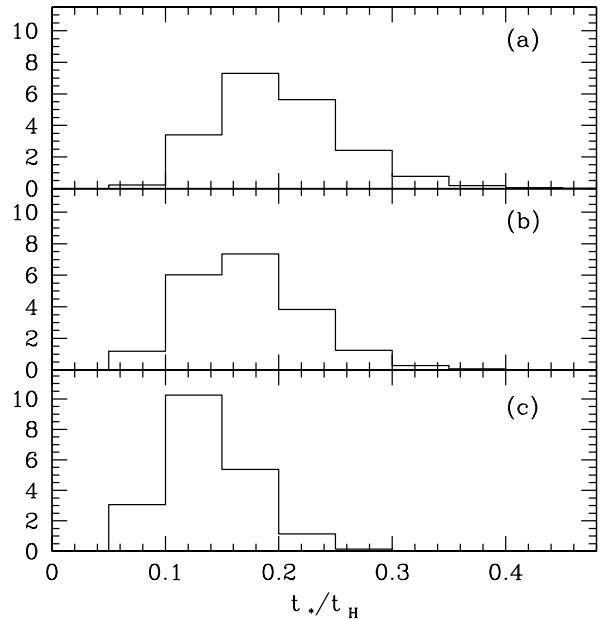}}
\caption{Histograms of the predicted distributions of the
star formation timescales (eq. \ref{tstar}) in units of the Hubble time
($\th$) at redshift 3 for the LBG population. The top, middle and bottom panels
correspond to the three cases of dust correction (a), (b) and (c)
defined in Section 2.1, respectively.
}
\end{figure}

Since LBGs are objects with high SFR at $z=3$, an important question is
how long can the high star formation activity be sustained. In our
model, we can easily estimate the star formation timescale
$\tstar$, defined as,
\beq \label{tscale} \label{tstar}
\tstar = {\Mgas \over \Mdot},
\eeq
where $\Mgas$ is the total gas mass in Equation (2) and $\Mdot$ is
the SFR. Note that in our empirical approach, the star
formation timescale for LBGs is independent of the parameter
$\md$, and is only a function of the dust-correction factor.

The predicted distributions of $\tstar$ in units of the Hubble time
$\th \approx 1.5h^{-1}\, {\rm Gyr}$ at redshift 3 are shown in Figure
6, where the three panels are again for the three cases of
dust corrections, respectively. This figure shows that,
as the dust correction increases, the predicted star formation
timescales become shorter on average due to increased intrinsic SFR.
In all cases, the star formation timescale is of the order of a few tenths of
the Hubble time, typically $\sim 0.2h^{-1}$\,Gyr. This timescale is 
somewhat longer than the typical
timescales of starbursts in the local universe, $\sim 0.1$Gyr, but
much shorter than the Hubble time. This
short star formation timescale relative to the Hubble time
implies that there must be more haloes hosting inactive `LBGs' 
than the observed number of LBGs; these haloes are not
currently forming stars actively so that they are not observable
as LBGs. The short duty cycle of LBGs has important
implications for the abundance of the host haloes, which we will discuss next.

\subsection{Constraints from the halo abundance}

As discussed in the previous subsection, the empirically
inferred star formation timescales for LBGs are only a few tenths
of the Hubble time at redshift 3. If we assume the lifetime of
a halo is typically a Hubble time, then the short
duty cycle means that there must be more haloes that
once hosted LBGs; most of these hosts are
dormant at $z=3$. Accounting for this duty
cycle, we can determine the comoving number density
of LBG haloes as a function of $\vc$ from the dust-corrected
luminosity function shown in Figure 2.  Figure 7(a) shows 
the required number density of LBG haloes as a function of $\vc$ for
the dust correction case (b); the solid, dotted and
dashed lines are for $\md=0.12, 0.06$ and 0.03, respectively.
The thick solid curve shows the halo abundance predicted by the
Press-Schechter formalism (Press \& Schechter 1974) in the
$\Lambda$CDM model (with $\Onow=0.3$, $\Omega_\Lambda=0.7$,
$h=0.7$, and with a shape parameter $\Gamma=0.2$ and 
a normalization $\sigma_8=1$ for the power spectrum).
 
In the above calculation, all the gas has been assumed to have collapsed 
into the disk and forming stars. However, one can imagine an alternative
scenario where initially only some gas collapsed into the disk; the
gas left in the halo then acts as a gas reservoir to prolong the star
formation timescale. As argued in \S2.1, the total fraction of gas mass
that can form stars must be smaller than the mass fraction in baryons,
$\fB \approx 0.12$. Fig. 7(b) shows the predicted number of haloes 
where we have assumed that all baryons are available to 
fuel the star formation.
 
\begin{figure}
\epsfysize=9.5cm
\centerline{\epsfbox{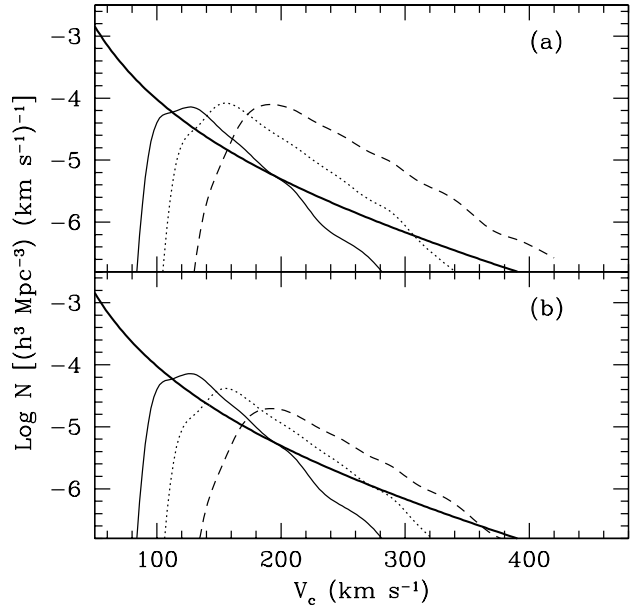}}
\caption{The inferred comoving number density of halos hosting LBGs as a 
function of the circular velocity $V_c$ for the 
dust correction case (b). The solid, dotted and
dashed lines correspond to $\md=0.12$, 0.06 and 0.03, respectively.
The thick solid curve is the predicted halo abundance by the
Press-Schecter formalism. 
(a) only accounts for duty cycle, while (b) accounts for additional gas
reservoir that may exist in LBG host haloes (see Section 3.3).
}
\end{figure}

Figure 7 strongly suggests that the mass ratio $\md$ cannot be much 
smaller than 0.1 if one considers the duty cycle of individual
LBGs but allows no gas reservoir. Otherwise, the required
halo number density will exceed the predicted number in the
$\Lambda$CDM model. Even if additional gas reservoir in a halo 
is considered, one still requires $\md \ga 0.06$.

 We have carried out the same calculation for a model
with $\Omega_{\rm m, 0}=1$, $\Omega_\Lambda=0$, $h=0.5$,
$\Gamma=0.2$ and $\sigma_8=0.6$. This model requires  
$m_{\rm g}\ga 0.1$ in order to predict sufficient number of 
haloes with $V_c>200\kms$. The required $m_{\rm g}$ is slightly larger
than that allowed by cosmic nucleosynthesis. 

\subsection{Correlation functions}

Based on the predicted distribution of the circular velocity $\vc$
above, we can calculate the correlation function for the LBG population
using the method suggested by Mo \& White (1996; 
see also Mo, Mao \& White 1999, hereafter MMW99).
The predicted correlation length $r_0$ as a
function of the median value of $\vc$ for the LBG
population (assuming varying $\md$)
is shown in Figure 8, where we have adopted
the dust correction case (b) (the result is very insensitive 
to the assumed dust correction).

As we have discussed \S3.1, the median value of $\vc$ for the LBG
population is in the range of 130-165$\kms$. From Figure 7, 
we conclude that the correlation length $r_0$ is in the range
2.6 to 3.4$\mpch$ for the $\Lambda$CDM model. This value is
smaller than the observed value obtained by Adelberger et al. 
(1998), but is consistent with the more recent determination
by Giavalisco \& Dickinson (2000) who found a
correlation length $(3.2 \pm 0.7)h^{-1}\Mpc$ 
(assuming the same flat cosmology as we are using here)
for a sample of LBGs brighter than $R=25.5$.
The predicted correlation length is also smaller than 
those given by many earlier studies, because of the smaller
halo circular velocities we predict for the LBG population.

\begin{figure}
\epsfysize=9.5cm
\centerline{\epsfbox{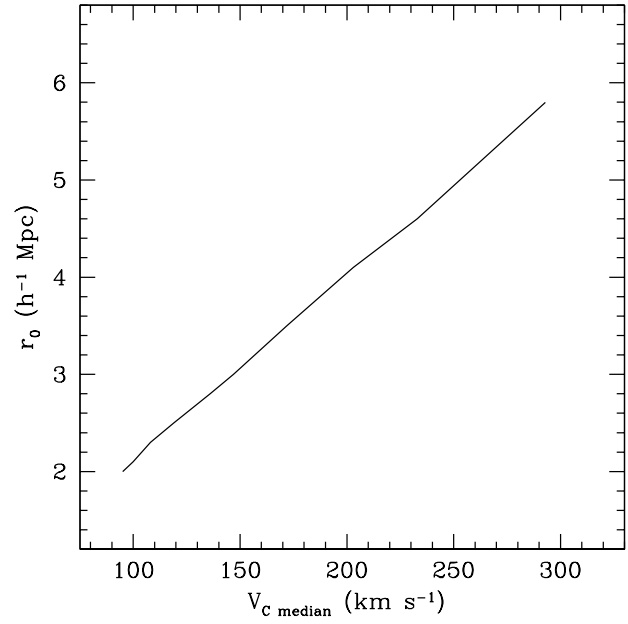}}
\caption{The predicted comoving correlation length $r_0$ as a
function of the median value of circular velocity $\vc$. The
dust correction is taken to be case (b).}
\end{figure}

 Using the model given in Mo \& White (1996), one can also predict  
the correlation function of the LBG descendants at 
the present time, assuming that mergers among LBGs are
not significant. The predicted correlation length is
about $6\,h^{-1}{\rm Mpc}$, comparable to that
of normal galaxies in the local universe.

\subsection{Stellar velocity dispersions}

\begin{figure}
\epsfysize=9.5cm
\centerline{\epsfbox{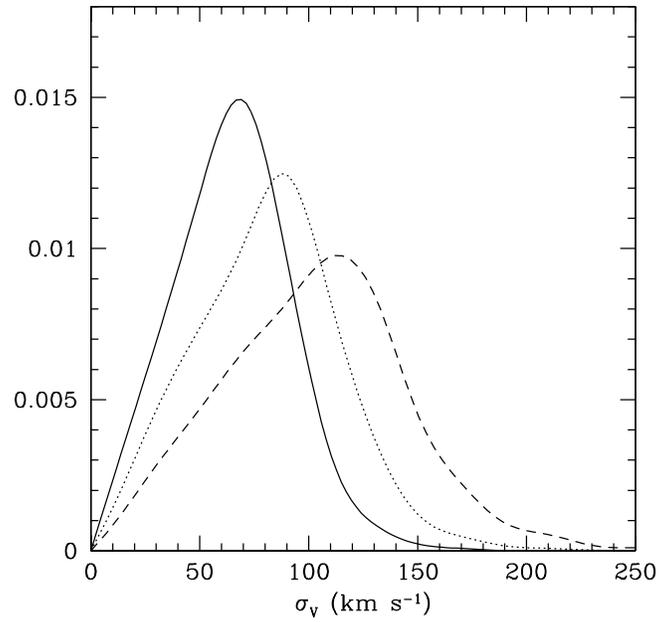}}
\caption{The predicted distribution of stellar velocity dispersions
for LBGs. The solid, dotted and dashed lines correspond to
$\md=0.12$, 0.06 and 0.03, respectively.
The dust correction is taken to be case (b).
} 
\end{figure}

Our models predict the circular velocities of the haloes
hosting LBGs, which are not directly observable. A
more observationally oriented quantity is 
the stellar velocity dispersion ($\sigmav$) weighted by the SFR in
the LBG population. As argued in MMW99,
the dispersions are in general
smaller than the underlying halo circular velocities as a result
of projection effects and the fact that the stars only sample the inner
rising part of the halo rotation curve. Following MMW99, 
we estimate $\sigmav$ for each LBG by
\beq
\sigmav^2 =
{1 \over 2\dot M_{\star}}\int_{0}^{\infty}V^2_{\rm rot}(r)
	\sin^2 i\,\Sigma_{\rm SFR}\,2\pi r\,{\rm d} r\,,
\eeq
where $\Sigma_{\rm SFR}$ is the SFR per unit area, $\Mdot$ is the total
SFR, $V_{\rm rot} (r)$ is the rotation speed at radius $r$, 
$i$ is the inclination and $i=0$ means a face-on disk.
We take $\sin i$ to be uniformly
distributed from 0 to 1. The rotation curve is assumed to be
produced by a dark matter halo with Navarro, Frenk \&
White (1997) profile; we take
the concentration parameter to be 10, the median
value found in numerical simulations (e.g. Jing 2000).

The predicted distribution for $\sigmav$ is shown in Figure 9 for the
dust correction of case (b). The solid, dotted
 and dashed lines are for $\md=0.12,
0.06$ and 0.03,  respectively. We see that
the median vulue of $\sigmav$ decreases with
increasing $\md$ because the predicted halo circular velocities
decrease as $\md$ increases (see Figure 3). As pointed out in
Section 3.1, a reasonable range of $\md$ is between 0.06 and 0.12, which
corresponds to a predicted median value for $\sigmav$ of about $70\kms$.

The stellar velocity dispersions of LBGs may be estimated
from the widths of nebular emission lines. A preliminary analysis of
about a dozen objects gives velocity dispersions in the range
$60-120\kms$, with a typical value of about $80\kms$ 
(see Pettini 2000). Our prediction is compatible with this
observational result. Clearly the observational sample so far is too 
small to draw a robust conclusion. We also point out that the 
correspondence between the emission-line widths
and the stellar velocity dispersions may not be
straightforward (see MMW99 for more discussions). Obviously,
a larger sample is needed to firmly test our prediction.

\section{Predictions for bright sub-mm Sources}

As we have shown in the previous section, 
haloes hosting LBGs are predicted to
have circular velocities in the range 
$100-300\kms$ with a median value of $\sim 150\kms$.
As one can see from Figure 7, the expected halo number density by the 
$\Lambda$CDM model in this range is comparable to the
required density. However, the $\Lambda$CDM model predicts
a substantial number of haloes with $\vc>300\kms$. A question
naturally arises: what are the observational manifestations of these even
larger haloes? Since the SFR is expected to be a steeply 
increasing function of $\vc$ within our theoretical 
framework, the SFRs in the central parts of 
these big haloes are expected to be much higher. 
As mentioned in Section 1, if the bright sub-mm sources 
observed by SCUBA indeed have redshifts $\sim 3$, their 
SFRs must be  $\sim 1000\,{\rm M_\odot\, \rm yr^{-1}}$,
much higher than the typical SFR for the UV-selected LBGs. 
It is therefore very tempting to make a connection between the 
bright sub-mm sources and the massive haloes expected 
in the CDM cosmogonies. Below we explore in detail the 
consequences of this connection. Throughout this section, we take
 the mass fraction of cold gas in 
dark haloes ($\md$) to be 0.1, which is within the plausible 
range discussed in Section 3.1.

 We identify bright sub-mm sources from Monte Carlo simulations
as follows. The halo circular velocities are drawn from the 
distribution predicted by the Press-Schecther formalism
for the $\Lambda$CDM model. Each halo is assigned 
a compactness parameter $\lambda_{\rm eff}$ (defined
in equation 5) drawn from a distribution
identical to that for the LBG population. The SFR
in each halo is derived as described in Section 2. 
We identify the bright sub-mm sources as systems which
have SFRs higher than that of the LBGs with the 
highest SFRs. Since the brightest UV-selected
LBG has $R=20.75$ ($\sim 1000 \my$) after applying
the dust correction suggested by Adelberger \& Steidel (2000),
we choose the bright sub-mm sources as galaxies with SFR 
$\ga 1000\my$. This selection procedure is obviously 
an oversimplified approximation. In reality, whether 
a source can be detected as a bright sub-mm source 
depends not only on the intrinsic SFR, but also on the
dust distribution around the galaxy. For example, 
a galaxy with high SFR may be more likely to be 
observed as a bright sub-mm source when it is more compact.
Unfortunately, theoretical models of dust emission
from star forming galaxies are still very uncertain 
and it is not yet realistic to adopt a selection criterion 
based on first principles (see, e.g., Granato et al. 2000). 
Because of this, our theoretical predictions are 
really only for galaxies 
with the highest SFRs. The connection to the 
bright sub-mm sources can be made only under the assumption that 
these sources are among the objects with the highest 
SFRs. With this assumption, many of our predictions 
(such as the circular-velocity distribution and correlation 
functions) are quite insensitive to the details of the 
selection criterion, because they are generic 
consequences of the high SFRs.

\begin{figure}
\epsfysize=9.5cm
\centerline{\epsfbox{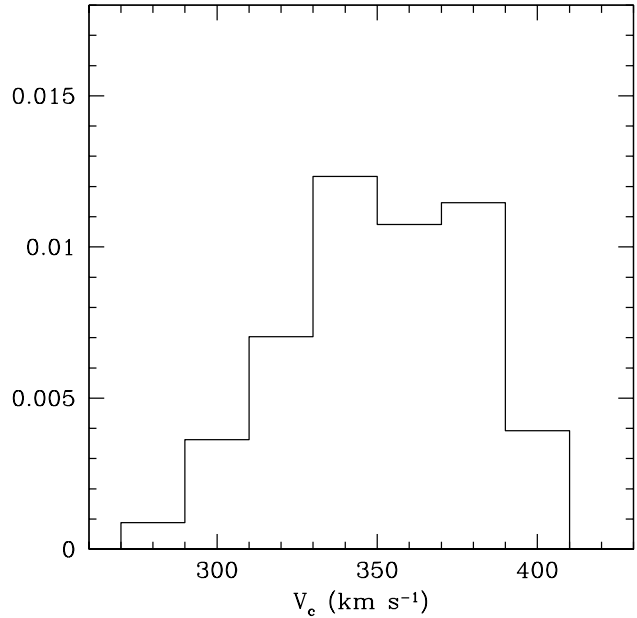}}
\caption{Predicted distribution of circular velocities $\vc$ for 
the haloes that host bright sub-mm sources.}
\end{figure}

\subsection{Circular velocities and sizes of the sub-mm sources}

\begin{figure}
\epsfysize=9.5cm
\centerline{\epsfbox{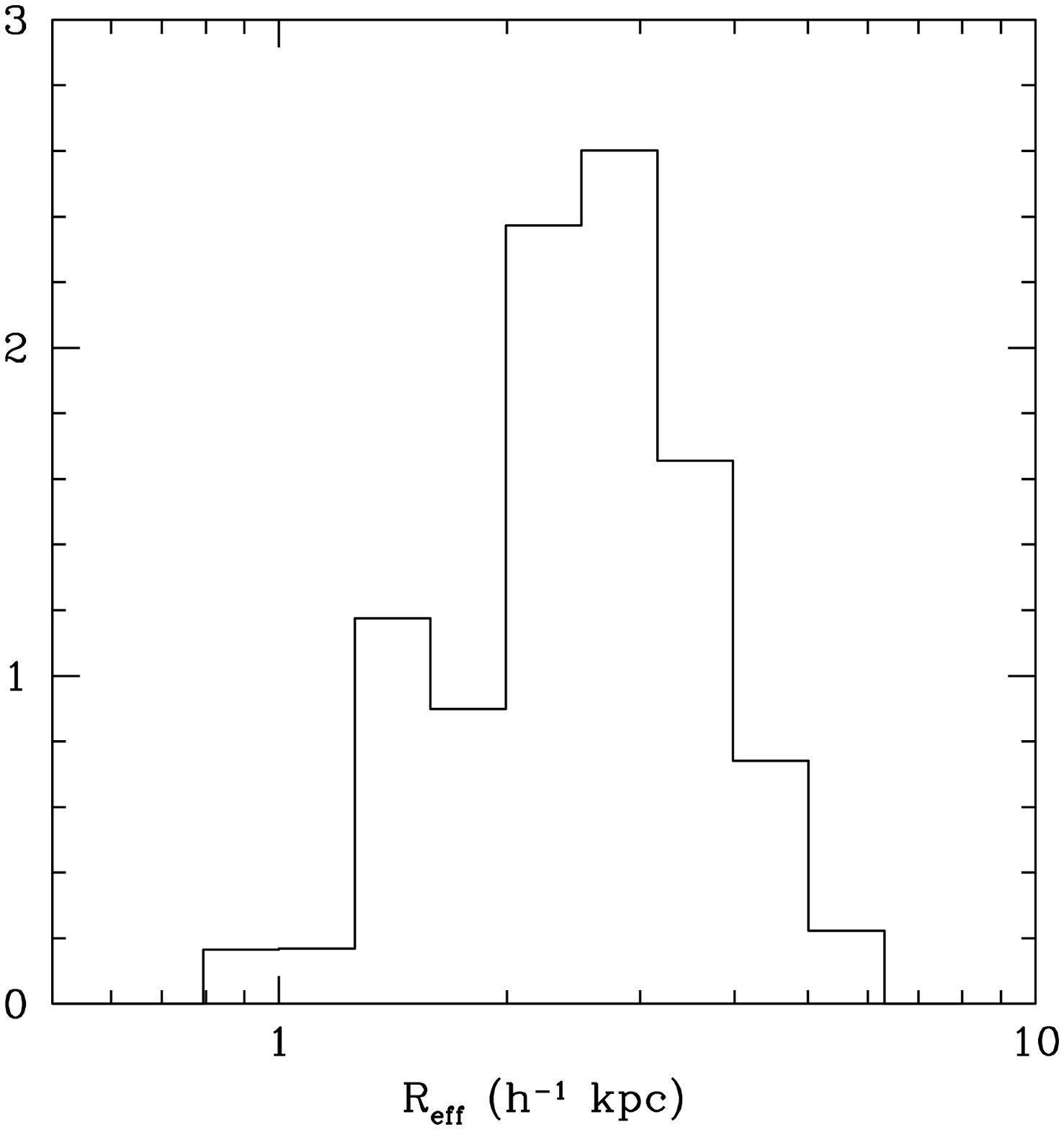}}
\caption{The predicted size distribution for SCUBA sources.}
\end{figure}

The predicted distribution of halo circular velocities for
the bright sub-mm sources is plotted in Figure 10.
The range of $\vc$ is from 270 to 410$\kms$ with
a median circular velocity of $\sim 350\kms$. Compared with the 
predictions for the LBGs (Section
3.1), the haloes of the sub-mm sources are predicted to
have much higher circular velocities. The predicted distribution
of effective (half-light) radius is shown in Figure 11. 
The median radius, $\sim 2.5\kpch$, is about twice 
as large as the observed median value for the LBG 
population (see Figure 1). 

\subsection{Star formation timescale}

\begin{figure}
\epsfysize=9.5cm
\centerline{\epsfbox{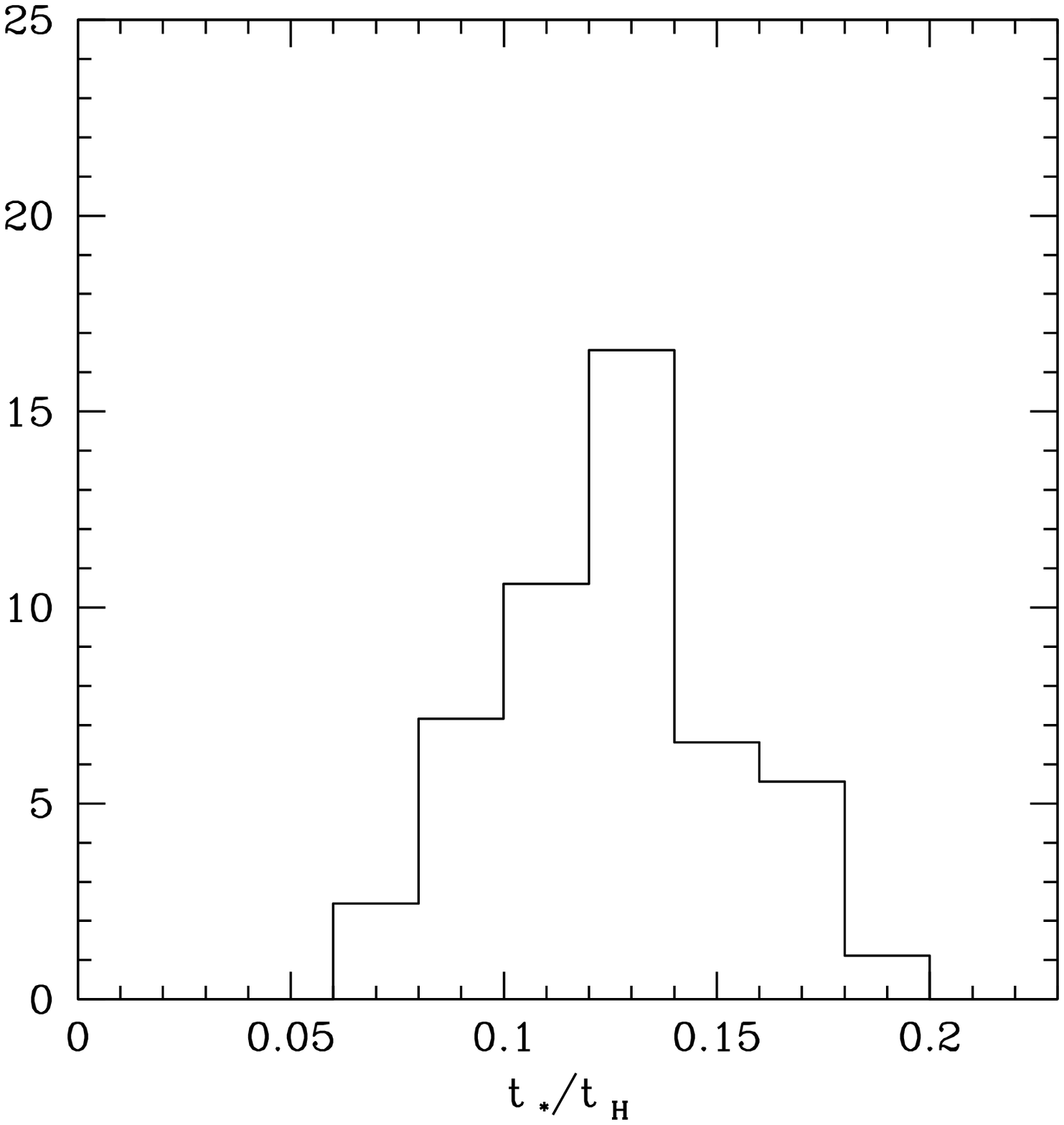 }}
\caption{The predicted distribution of star formation timescales
for bright sub-mm sources relative to the Hubble time ($\th$).}
\end{figure}   

Because the bright sub-mm sources all have high SFRs, 
it is interesting to examine their typical star formation 
timescale. If we define the star formation time in the same 
way as given by Equation (6),
the distribution of $\tstar$ for
the sub-mm sources looks like that shown in Figure 12. 
The median value of the distribution is about 10
per cent of the Hubble time at redshift 3, somewhat smaller
than that predicted for the LBGs (Section 3.2). 

Taking into account this duty cycle, the comoving number density 
of such objects at $z\sim 3$ is predicted to be 
about $10^{-4} h^3 \mpc^{-3}$ in the $\Lambda$CDM model, which is
comparable to the observational result of the sub-mm sources with
$S_{850\mu m} \ga 10\rm mJy$ assuming their redshifts $z\sim 3$
(Dunlop 2000).  
It must be pointed out that the predicted number density
is quite sensitive to the SFR threshold adopted in 
the selection of the sub-mm sources, because the massive haloes
are already on the tail of the halo mass function
at $z=3$. However, the prediction that the haloes of bright 
sub-mm sources are massive is generic because, 
according to the empirical star formation prescription,
the SFR strongly depends on halo circular velocity 
($\propto V_c^{3.4}$) but only weakly on the compactness
parameter ($\propto \lambda_{\rm eff}^{-0.8}$).

\subsection{The correlation functions}

Using the model of Mo \& White (1996), we can calculate 
the correlation function for the strong sub-mm sources
at $z\sim 3$. The inferred correlation length is 6.8$\mpch$, 
about a factor of two larger than that predicted for 
the LBG population. Thus, bright sub-mm sources, such as the ones 
observed by SCUBA, should be very strongly clustered in space.
The corresponding correlation length for the descendants
of these sources at the present time, which can be calculated 
using the method described in Mo \& White (1996), is about $10\mpch$. 
This correlation length is larger than that for present-day 
field galaxies, but is comparable to that for massive elliptical 
galaxies. This suggests that the strong sub-mm sources 
may be the progenitors of massive elliptical galaxies
observed in clusters of galaxies today. With the same model we 
can also estimate the cross-correlation function between 
the LBG population and the sub-mm sources. We find a 
correlation length of about $\sim 4.5\mpch$ for the model
in consideration. 

 It should be emphasized here that although the predicted 
number density of the sub-mm sources is sensitive to the 
selection criterion, the predicted correlation lengths are not.
This is because while the number density of massive haloes
at $z\sim 3$ depends exponentially on the halo circular velocity,
the correlation amplitudes have a much weaker dependence on 
the halo circular velocity.
Thus, the predicted correlation lengths are robust 
predictions which can be used to test the model.
 
\subsection{The stellar velocity dispersions}

\begin{figure}
\epsfysize=9.5cm
\centerline{\epsfbox{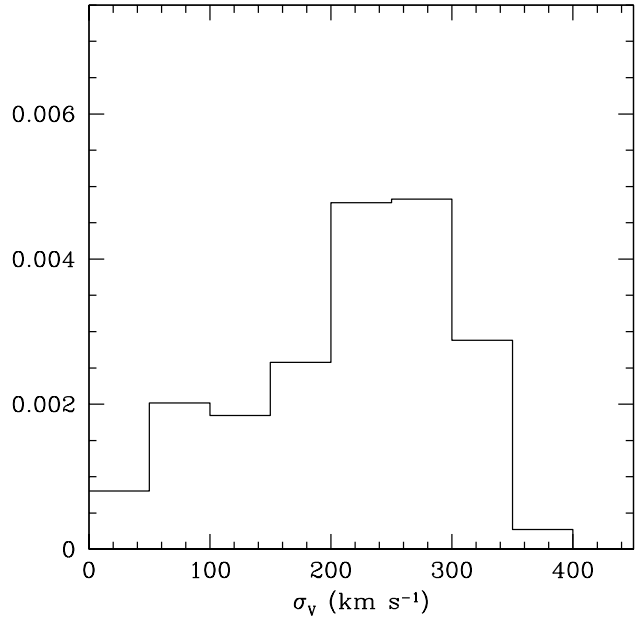}}
\caption{The predicted distribution of the stellar velocity
dispersions for bright sub-mm sources.}
\end{figure}

As for the LBG population, we also calculate the 
SFR-weighted velocity dispersions for the sub-mm
sources. The expected distribution is displayed in Figure 13. 
The distribution is quite broad with a median 
of $\sim 200\kms$, which 
is much larger than that predicted for the LBG population.
This larger stellar velocity is a direct consequence of
the fact that the host haloes for SCUBA sources are 
predicted to be much more massive.
This prediction is also fairly insensitive to the selection 
criterion. 

\subsection{The contribution to the 850$\mu$m background}

As we have discussed, bright sub-mm sources have 
high star formation rate and suffer severe dust 
obscuration. Most UV and optical
radiation is absorbed by dust and re-radiated in the FIR/sub-mm wavelength.
It is therefore important for us to estimate the contribution
of our model sources  to the sub-mm background. 
According to Scoville \& Young (1983) and Thronson \& Telesco  
(1986), the FIR luminosity is related to the SFR by
\beq \label{FIR}
{\rm SFR} \sim 1.5 \times 10^{-10} 
{ L_{\rm FIR} \over L_\odot}
M_\odot\rm  yr^{-1}.
\eeq 
Using our Monte Carlo sample, we can estimate the total contribution
of massive galaxies to the 850$\mu$m background by integrating the
intensities from redshift 2.5 to 3.5 by
\beq
\rho_{850\rm \mu m}= \int_{2.5}^{3.5} dz {dV_{\rm com} \over dz}
\int \,d L_\nu\, n(L_\nu, z) \, {L_\nu \over 4\pi d_L^2}\,,
~~~~\nu={c(1+z)\over {\rm 850\mu m}}\,,
\eeq
where $L_\nu (\vc, \lambdaeff)$
is the monochromatic luminosity at frequency $\nu$ of an individual 
sub-mm source, $n(z)$ is the comoving number density of sub-mm sources as a
function of redshift $z$, $dV_{\rm com}$ is the differential comoving
volume from $z$ to $z+dz$ and $d_L(z)$ is the luminosity distance
for redshift $z$. We assume the comoving number density for sub-mm
sources to be constant within the redshift range $z=2.5-3.5$.
For given SFR, the total FIR luminosity can be estimated using
Equation (8). However, to convert it to the monochromatic
luminosity (and flux), we
need detailed information about the source spectral energy distribution
which depends on the uncertain dust temperature. We estimate this by
using Equation (7a) in  Adelberger \& Steidel (2000) which connects the
monochromatic luminosity to the dust bolometric luminosity:
$L_{\rm bol,dust}=\nu L_\nu {\cal K}_{850}(z)$;
we take ${\cal K}_{850}=10$ and $L_{\rm bol,dust}\approx 1.47 L_{\rm FIR}$
(see Adelberger \& Steidel 2000 for details).

Carrying out the procedure outlined above, 
we find that the inferred contribution of the sub-mm sources
to the 850$\mu$m background is $1.2 \times 10^{4} h^2 \rm
mJy/deg^2$. Similarly, for the LBGs, we find a contribution of
$1.9 \times 10^{4} h^2 \rm mJy /deg^2$. The contribution of LBGs is
about the same as that of the sub-mm sources
because their larger number density compensates 
their lower SFRs. Both the contributions of the sub-mm sources and LBGs
are about 15 per cent of the
observed background, $\approx 4.4 \times 10^{4}\rm mJy /deg^2$ 
(Fixsen et al. 1998). The total SFR density for sub-mm sources and LBGs are
0.096, and 0.15$\my h^{3} \mpc^{-3}$. These estimates are consistent
with previous studies (Blain et al. 1999). We caution, however,
that this prediction for sub-mm sources,
unlike all the other predictions in this section,
depends quite sensitively on how we select sub-mm sources in the
Monte Carlo simulations.

\section{Summary and Discussion}
\label{sec_discussion}

In this paper we have adopted an empirical approach to investigate the
physical properties of LBG population. We have shown that
the observed sizes and star formation rates of LBGs, combined with the
Kennicutt star-formation relation (Kennicutt 1998), directly constrain the
circular velocity of the host haloes and the concentration
of the observed galaxies within them.
For the adopted $\Lambda$CDM model, we find that observations
require the circular velocities of LBG host haloes 
to be in the range $100-300\kms$, and with a median
of $\sim 150\kms$. This empirically determined range of circular
velocities is different from those found in previous studies,
because we find that star formation in LBGs has a relatively
short duty cycle; a more detailed comparison is given in
in Section 5.1. 
The star formation in our model haloes is quite efficient;
almost all the gas in them can form stars on a time scale
$\sim 3\times 10^8\,{\rm yr}$, much shorter than the Hubble time at
$z=3$. This means that there must be more haloes that can  host
LBGs than we actually observed; this has important consequences for
the abundance of the descendants of LBGs, a point which we return to 
in Section 5.2. 
The predicted comoving correlation length 
of LBGs is $\sim 3\, h^{-1}{\rm Mpc}$, and 
the predicted velocity dispersion of their stellar 
contents is typically $70 {\rm \,km\,s^{-1}}$. Both these predictions are
compatiable with observations (Giavalisco \& Dickinson 2000;
Pettini et al. 1998a; Pettini 2000).

 The same prescription applied to larger haloes predicted in 
the CDM cosmogony predicts the existence of galaxies 
with star formation rates $\sim 1000\my$ at redshift $z\sim 3$.
We identify these galaxies to be the bright sub-mm sources
detected by SCUBA. The model predicts that the host haloes
of these sub-mm sources are massive, with typical 
circular velocity $\sim 350\kms$ and typical stellar velocity
dispersion $\sim 200\kms$. Both the halo circular velocity and velocity
dispersion are about a factor of $\sim 2$ larger than those for the LBGs.
The typical star formation timescale in these systems
is about 10 per cent of the Hubble time at redshift 3, and
the comoving number density of galaxies (in their duty 
cycle) is $10^{-5}$--$10^{-4}h^3 {\rm Mpc^{-3}}$.

Since the LBGs and the bright sub-mm sources are both highly dust-obscured
SFR galaxies, most of their energies are absorbed by dust and
re-radiated in sub-mm sources, we estimate that 
they make comparable contributions (about 35 per cent in total) 
to the sub-mm background.

\subsection{Comparisons with previous studies\label{sec_comparison}}

In our model, the predicted circular velocities of the host haloes 
are smaller than those found in most previous analytical and
numerical studies (see Section 1 for more complete references).
For example, by matching the number density of massive haloes
with the number density of LBGs, the haloes selected as LBG hosts
have circular velocities $\ga 200\kms$ (MMW99; Baugh et al. 1998).
The difference arises from several factors. First, our approach is
more empirically based since we directly use the
observed size and SFR distributions. For given $\md$,
the inferred $\vc$ only depends on two assumptions, i.e.,
the star formation occurs in an exponential disk 
and the Kennicutt (1998) SFR relation applies at $z=3$. The first assumption
is consistent with the surface photometry of LBGs
(Giavalisco et al. 1996; Lowenthal et al. 1997). The second
assumption is plausible, since it is satisfied by local starburst galaxies.
In any case, this assumption, in one form or the other,
has been adopted in all studies. We therefore believe
that our estimate of $\vc$ relies less on theoretical modelling
and should be reasonably robust. Second, the
dust-correction we favor (Adelberger \& Steidel 2000) is about a factor
of two or three higher than that used by previous studies; 
this implies larger intrinsic SFRs for the LBG population. 
Third, due to the increased intrinsic 
SFRs, all gas in the host haloes of LBGs
is typically consumed within a few tens of
the Hubble time at $z=3$. This means that the duty cycle of the LBG
phase must be considered, an effect largely ignored in previous studies.
A direct consequence of the duty cycle is
that the number densities of haloes that can host LBGs must be larger
than that of the observed LBGs (see \S5.2).
To satisfy this increased number density of host
haloes, their circular velocities have to be correspondingly
reduced. The reduced $\vc$ leads to a correlation length
$r_0\sim 3\, h^{-1}{\rm Mpc}$, which is smaller than 
that given by previous studies (e.g. Adelberger et al. 1998)
but consistent with the recent determination
by Giavalisco \& Dickinson (2000).

\subsection{Descendants of LBGs and sub-mm sources \label{sec_descendant}}

The observed number density of LBGs down to $R=25.5$ is $\sim 2\times 
10^{-3}h^3{\rm Mpc^{-3}}$ (Adelberger et al. 1998). As
the star formation time scale is only about fifteen per cent of the Hubble
time at $z=3$ in our model (see the middle panel in Figure 6), 
the haloes of possible LBG hosts must be about a factor of $\sim 6$
more abundant than the observed LBGs. 
The descendants of their stellar components would still have 
number density of about $\sim 10^{-2} h^3 \mpc^{-3}$, if merging is
not dominant, larger than the number density of
local $L_\star$ galaxies. It is therefore possible that
some of the LBGs have survived mergers to become 
the progenitors of the stellar bulges in disk galaxies 
we see today. The sizes and stellar velocity dispersions both match
the values for the spheroidal components of disk
galaxies such as the Milky Way. Further the (comoving)
correlation length predicted for their descendants is also similar to the 
present-day normal galaxies.
Since the star formation time scale of LBGs is $\sim 0.2h^{-1}\Gyr$
with typical SFR of about $50\my$, the total stellar mass formed 
is $\sim 10^{10}M_\odot$, comparable to the mass of stellar bulges.

Our model predicts that the bright sub-mm sources are 
hosted by more massive haloes with $\vc \sim 350\kms$, and their
typical stellar velocity dispersion is $\sim 200\kms$.
These kinematical properties are very similar to those found in
elliptical galaxies, therefore it is natural to make a connection between
the bright sub-mm sources and the present-day giant ellipticals
(see also Dunlop 2000). The predicted number density of these objects is
about $10^{-4} h^3\mpc^{-3}$ (see \S 4.1). Because the SFRs in these objects
are very high ($\sim 10^3\my$), the star formation proceeds rapidly, and
all gas is consumed in
only about ten per cent of the Hubble time at $z=3$. The required
duty cycle implies that the number density of the remnants of sub-mm
sources, if they survive today,
can be as high as $\sim 10^{-3}h^3\mpc^{-3}$. This is within a factor
of few of the observed 
number density of giant ellipticals in the local universe. 
The total stellar mass formed, $\sim 10^{11} M_\odot$,
is also comparable to the stellar mass seen in giant ellipticals. 

An important question is what
triggers the massive star formation in the strong sub-mm sources? In the local
universe, the most extreme starbursts are found in dusty ultraluminous
infrared galaxies, a local analogue of the sub-mm sources. 
Nearly all these galaxies
occur in merging galaxies (see Sanders \& Mirabel 1996 for a
review). In Figure 14 we show the merging probability for haloes
as a function of the circular velocity $\vc$ at redshift 3 for the
$\Lambda$CDM model based on the extended Press-Schechter
formalism (Bond et al. 1991; Bower 1991; Lacey \& Cole 1993).
Interestingly, one sees that the maximum merging rate occurs for
$\vc \approx 350 \kms$, nearly identical to 
the predicted median value of $\vc$ for bright sub-mm sources.
This suggests that massive
starbursts may be triggered by major mergers, as seen in the 
local universe. Similar to the local
ultraluminous IRAS galaxies, some fraction of the energy in them may be
contributed by a central AGN fueled by the accreted gas (Genzel et
al. 1998). 

\begin{figure}
\epsfysize=9.5cm
\centerline{\epsfbox{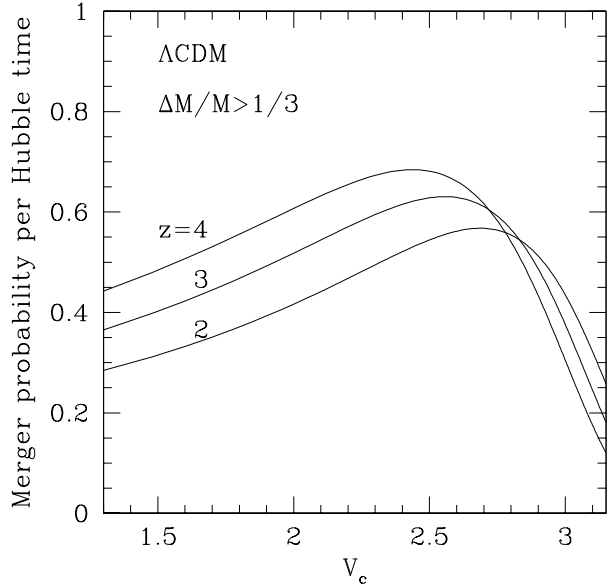}}
\caption{The halo merging probabilities as a function of $\vc$.}
\end{figure}

To conclude, we have modelled the LBGs and sub-mm sources using
an empirical approach. This approach makes a number of robust predictions
about the correlation, sizes and kinematical properties of bright 
sub-mm sources. One interesting consequence of the model
is the existence of a population of inactive 
haloes at $z\sim 3$. The number density of these
dormant haloes is a factor of 6-10 higher than the observed 
density of LBGs and SCUBA sources. The massive ones are
expected to contain large amount of relatively old stars
and can already be seen in the $K$-band. It would be very 
interesting to explore further the observational signatures of 
these objects.

\section*{Acknowledgement}

The authors thank Simon White for carefully reading 
the manuscript and useful suggestions.
CS acknowledges the financial support of MPG for a visit to
MPA. SM is grateful to MPA for hospitalities during a visit.
This project is partly supported by the Chinese National Natural
Foundation, the WKC foundation and the NKBRSF G1999075406.

\end{document}